# Improvement of Memory Characteristics of Charge-Trapping Memory Device by Using HfO$_2$/Al$_2$O$_3$ Laminate

Yifan Hu[1*]

[1]Sungkyunkwan University, Department of Electronics and Computer Science, Republic of Korea
Email: ivanhu@g.skku.edu

**Abstract:** Current portable memory device relies heavily on flash memory technology for its implementation. New generation of non-volatile memory is likely to replace floating gates, charge-trapping memory currently still suffering from inadequate retention performance and slow programming speeds as well as small memory window. In contrast, the use of Al$_2$O$_3$/HfO$_2$ stacked high-k materials has substantially increased the memory window of memory devices, with a 63% increase relative to pure HfO$_2$ materials. The current goal of up to 80% retention characteristics over a ten-year period has been achieved by adjusting the deposition ratio or dopant material. Process conditions can continue to be investigated in the future to reduce charge loss while maintaining the increased memory window.

## 1. INTRODUCTION

Memory devices have played a very important role in the information industry in recent years with almost all industries including them in their semiconductor memory systems. However, as Moore's Law has become irrelevant, the demand for both performance and size has increased, with the desire to further miniaturize devices while improving memory performance without losing data in the event of a power failure Memory devices have played a very important role in the information industry in recent years[1,2]. There are also new types of non-volatile memory (NVM) devices, such as resistive memory (RRAM), magnetic memory (MRAM) [3] and phase change memory (PCM) [4], as shown in *Fig*.1. However, most of these memory devices suffer from small memory windows, high energy consumption, slow programming speed, and poor retention performance. To solve these problems, researchers have proposed a new charge-trapping memory device (CTM) [5]and applied high-k materials to it.

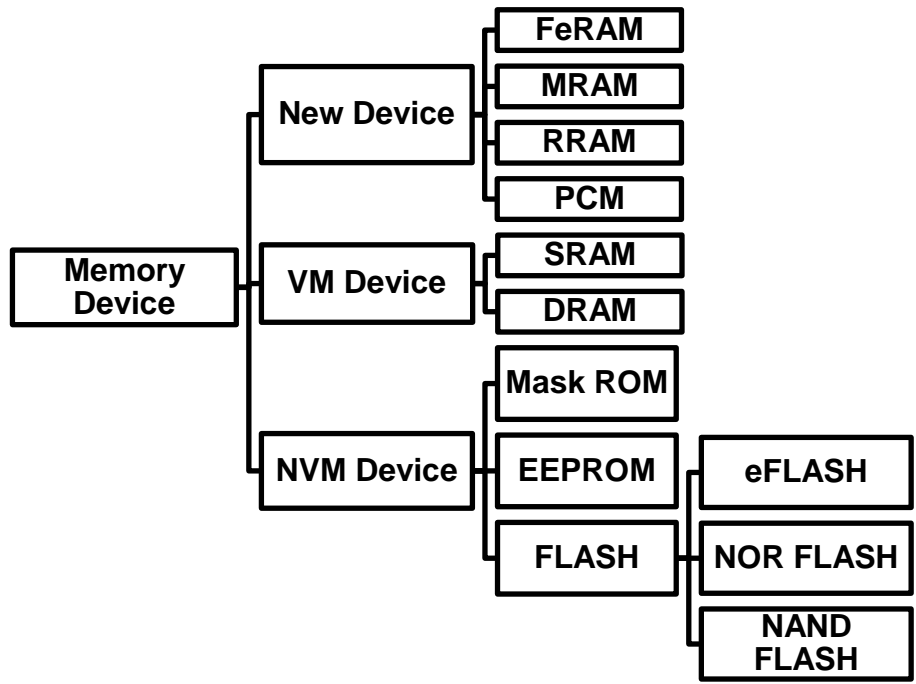

**Fig. 1.** Classification of memory devices

The concept of charge-trapping memory devices was introduced by Newman and Wegener in 1968[6], and the structure of Silicon-Oxide-Nitride-Oxide-Silicon (SONOS) was refined and proposed by Chen et al.[7] and the introduction of high relative dielectric constant materials into charge-trapping memory devices was proposed by Sangmoo Choi et al[8]. They used high-k $Al_2O_3$/$HfO_2$ laminate to replace $SiO_2$ as the blocking layer in the device, which improves the retention characteristics and reduces the programming and erase voltages of the device, reduces the energy consumption of the device, and also greatly reduces the leakage current, improves the device's stability while in use. Since then, researchers have been investigating the feasibility of using $HfO_2$ and $Al_2O_3$ as charge-trap layers and improving their performance. In 2010 Tackhwi Lee et al.[9] used $NH_3$ during rapid thermal annealing(RTA) of the device to form a HfON trapping layer, resulting in a charge-trapping layer with a very high dielectric constant, which increased the electric field in the tunneling oxide layer, resulting in faster programming and erasing speed.

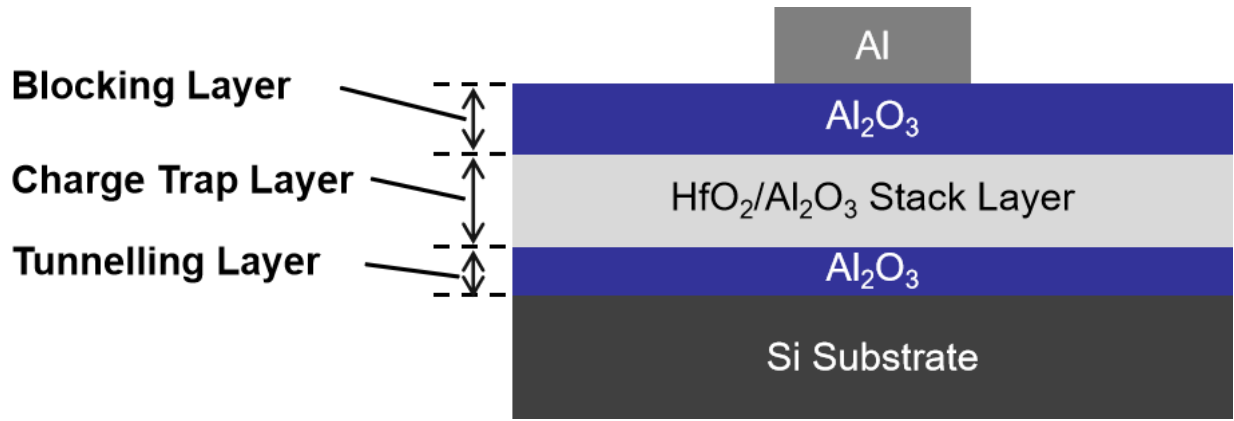

**Fig. 2.** $Al_2O_3/HfO_2$ based charge-trapping NVM Structure

The current NVM structure based on the $Al_2O_3/HfO_2$ stack is shown in *Fig 2*, high-k materials are required to achieve large memory windows, while large bandgap materials are required to achieve excellent retention properties. It is difficult to combine many good properties, so researchers have improved the performance of charge-trapping memory devices by modifying the structure of the trapping layer, doping the charge-trapping layer with atoms, or changing the process conditions during preparation.

## 2. METHOD FOR EXPERIMENT

Thin film preparation techniques play a very important role in the performance enhancement of charge-trapping layers. The parameters in the thin film preparation process can have different effects on the homogeneity, denseness, and crystallinity of the film, therefore on the performance of the capture layer. Atomic Layer Deposition (ALD) is currently the main thin film preparation instrument for MOHOS devices because it produces uniformly dense films and does not require complex operating systems to prepare composite film stacks [10]. Also, ALD enables doping of the material, usually by splitting a large ALD cycle into two ALD half-cycles. The cycle usually starts with the deposition of the substrate material followed by the deposition of the dopant material. The desired thickness is achieved by repeating the large cycle several times [11]. In this process, the dopant concentration is adjusted by varying the proportion of dopant material circulating in the large cycle, and the operating parameters of the ALD process are usually tuned to optimize the performance of the charge memory layer. Typically, after cleaning the wafer, ALD system is used to deposit trapping layer blocking layer and a tunneling layer, then use rapid thermal annealing (RTA) treatment to stabilize the lattice structure and improve the quality of the film [12], and finally make the electrodes using thermal evaporation deposition, the process is shown in *Fig.3*.

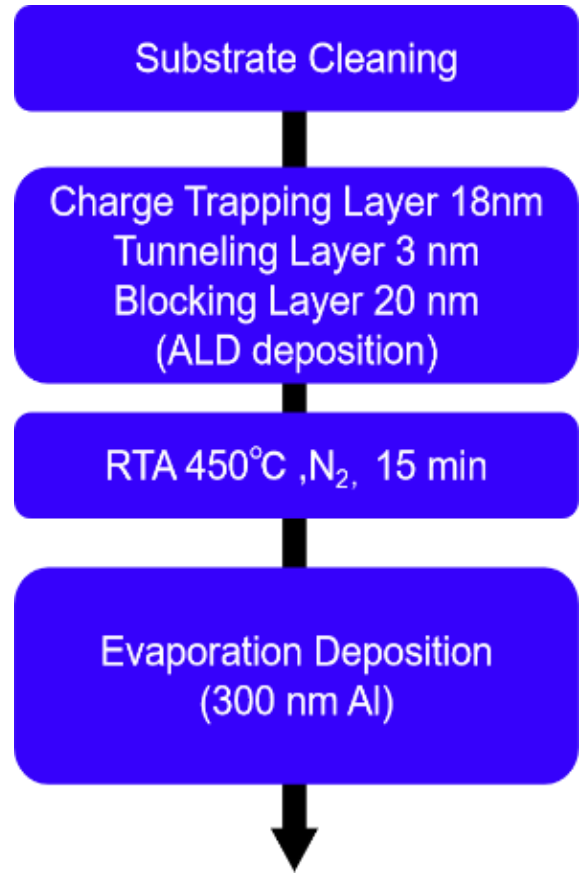

**Fig. 3.** Experimental flow chart for charge-trapping memory preparation

## 3. RESULTS AND DISCUSSION

### 3.1 Memory Window

Memory window is an important property which is used to characterize the memory capacity of a memory device that reflects the degree of distinction between different data storage states of '0' and '1'. The memory window is usually expressed in terms of the difference in voltage between the two flat band voltages of the C-V bidirectional scan ($\Delta V_{FB}$). Memory window affects the judgement of the memory state so researchers have increased the memory window in many ways.

As shown in *Fig.4*, the memory window of the device is changed by varying the ALD deposition temperature. As the growth temperature increases, the desorption of $HfO_2$ and $Al_2O_3$ increases during the growth process, thus destroying the crystal structure in contact with the Si interface and generating more defects and generating interfacial traps. changing the temperature of the ALD has a very significant effect on the memory window and the interfacial charge density of the device of two $1.31\times10^{13}$ cm$^{-2}$ and $2.3\times10^{12}$ cm$^{-2}$ respectively, which are 32 and 6 times higher than those at 100 °C [13].

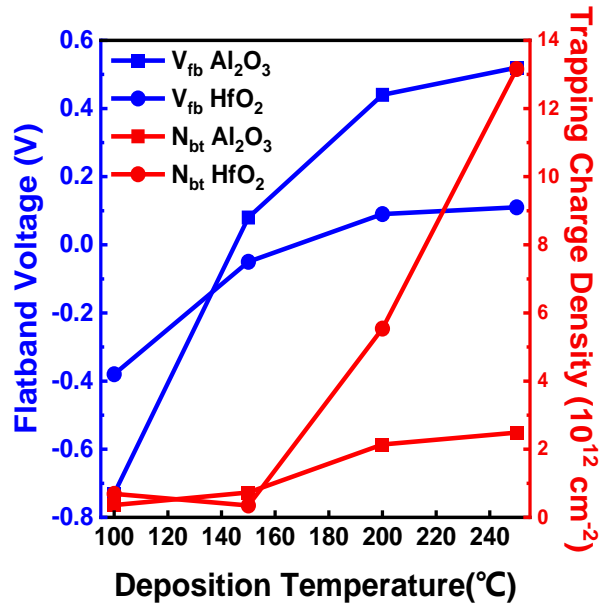

**Fig. 4.** Flat-band voltage and boundary trap density as a function of temperature

Dencho Spassov et al. [14] had conducted radiation treatment of on HfO$_2$/Al$_2$O$_3$-based NVM using γ-radiation by using Co-60 in 2021 to improve the memory window of the device. As shown in *Fig.5*, the results showed that γ-radiation enhanced the memory properties of the samples annealed in O$_2$ ambience in rapid thermal process (RTP). The rapid thermal processed samples and as-deposited samples showed improvement in storage capacity by 58.82% and 305%, respectively but only the devices annealed in N$_2$ ambience received severe damage in storage capacity after radiation because of the formation of HfON, which suppressed the charge vacancies caused by radiation. In contrast, γ radiation enhanced the charge-trapping in the native and oxygen annealed HfO$_2$/Al$_2$O$_3$ multilayer.

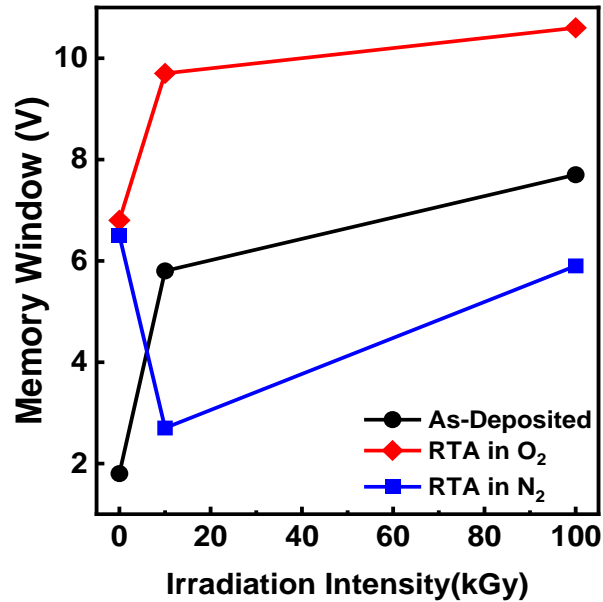

**Fig. 5.** Radiation exposure on memory windows under different thermal annealing conditions

In 2022，Dencho Spassov et al. [15] optimized device performance by changing the Hf:Al ratio. They deposited ultra-thin films with different ALD ratio and put them into RTA under $O_2$ atmosphere, which showed that the charge-trapping characteristics under RTA were changed a lot. The device which was deposited at Hf:Al ratio of 20:5 had reduced charge-trapping characteristics after RTA while the sweep voltage is over 30 V. The Fig.6 showed that the trapped net positive charge decreased and trapped net negative charge increased, while the Hf:Al ratio is 4:1. The memory window increased after RTA. In contrast, it was found that the higher the number of layers, worse is the memory properties because of thermal diffusion during RTA in too many stacked layers, thermal diffusion occurs at RTA which affects the ability of the captive layer to bind the charge.

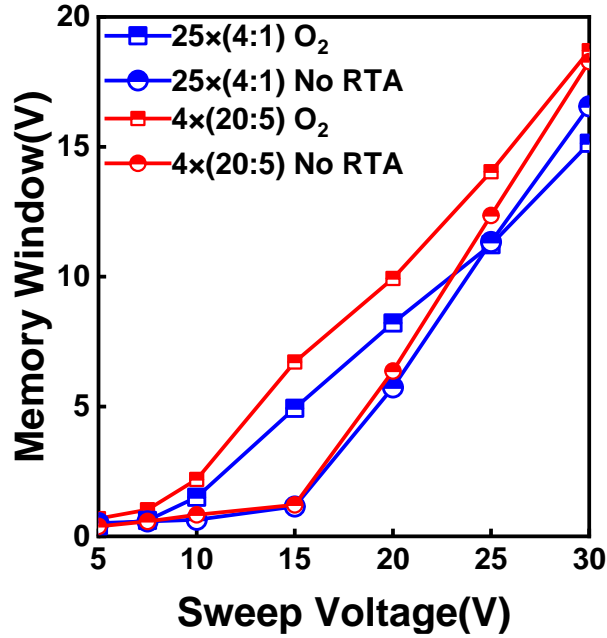

**Fig. 6.** Memory windows with different cycle ratios under RTA.

In addition to changing the trapping layer, Yoo, Jinhyuk et al. in 2019 attempted to optimize the memory window of the device by changing the content of Al in trapping layer[16]. In recent years, the materials $Al_2O_3$ and $HfO_2$ have been widely employed in memory systems, each with its own set of benefits. Their dielectric constants are in the range of 25. $Al_2O_3$ and $HfO_2$ are far less expensive to make than other high dielectric materials and have bandwidths in excess of 5 eV. Although $HfO_2$ has a high trap density, the defect energy level is not as high as $Al_2O_3$ and therefore has a weak ability to hold electrons, whereas $Al_2O_3$ has a deep trap energy level but is fast to program. Due to its high dielectric constant, $Al_2O_3$ can act as a barrier and tunneling layer to prevent charge leakage. When the Hf:Al ratio is 3:1 and scan voltage of 10 V is applied, as shown in *Fig.7*, the device achieves a memory window of 5.22 V, whereas when the Hf:Al ratio is 1:1, the memory window drops to 3.39 V. The device performance can be optimized by reasonably adjusting the ratio of the two materials.

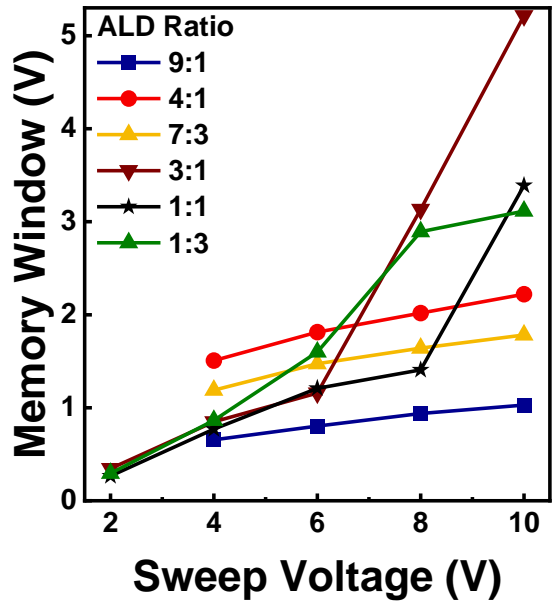

**Fig. 7.** Memory window for devices with different ALD $Al_2O_3$/$HfO_2$ cycle ratios.

### 3.2. Working stability performance

Dencho Spassov et al. [15] used the method by changing tunnelling layers to optimize the memory characteristics of the device, as shown in *Fig.8*. When the tunneling layer was 3.5 nm, the preservation characteristics were 16.7% higher after $10^6$ s of applied voltage than when the tunneling layer was 2.4 nm. Increase in the thickness of the tunnel layer can slow down charge loss, but at the same time it increases the operating voltage of the device.

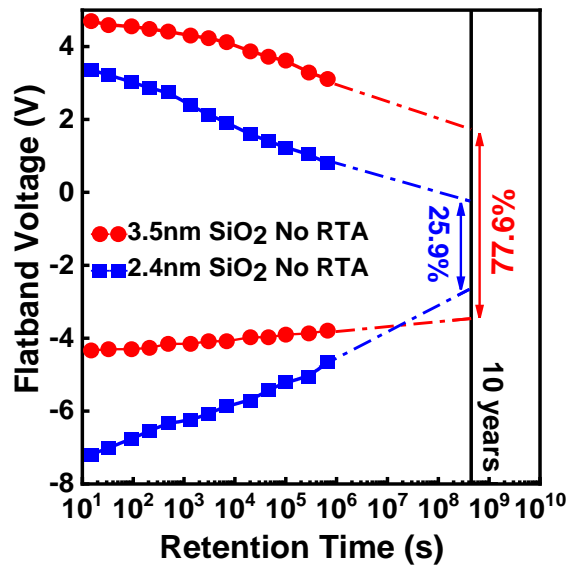

**Fig. 8.** Different thicknesses of tunnelling layers on retention performance

Researchers have also attempted to improve the retention property by varying the process conditions. Hou et al [17] compared two samples S1 (450 °C, 15 s) and S2 (1000 °C, 60 s) using different conditions of RTA of HfAlO samples at ±10 V P/E voltage and 1 s pulse width. The results show that the S1 sample maintains a memory window of 2.17 V after a retention time of $10^4$ s, corresponding to 69 % of the initial memory charge, while the S2 sample achieves a memory window of 3.14 V, corresponding to 74.76 % of the initial memory charge. The doping of $Al_2O_3$ in $HfO_2$ causes deep energy traps in the trapping layer, while high temperature annealing causes excellent doping of $Al_2O_3$ and $HfO_2$ in the charge-trapping layer to inhibit the charge from leaving the traps, and faster charge loss within the trapping layer due to the higher potential barrier of the holes. Also, the temperature and RTA treatment time can effectively improve the P/E speed time, by improving the P/E speed can effectively reduce the working time and reduce the power consumption. Hou et al. prepared $Al_2O_3/(HfO_2)_{0.9}(Al_2O_3)_{0.1}/Al_2O_3$ structure device and did RTA treatment on these devices at different temperatures and time conditions[17]. By studying the $\Delta V_{FB}$ of the C-V curve at 1 MHz, the P/E speeds of S1 and S2 were characterized. As shown in *Fig.9(b)*, the results show that S2 can obtain a $\Delta V_{FB}$ of 2.8V when programmed at +10V for $10^{-6}$ s, at this stage S1 can only obtain $\Delta V_{FB}$ of 1.9V. The programming speed of S1 is improved when P/E exceeds 0.1 ms, showing that high temperature and long RTA improves the programming and erasing speed of the device, but has no significant effect on the charge trapping capability of the device as the $\Delta V_{FB}$ of S1 and S2 does not change significantly when the P/E time reaches $10^0$ s.

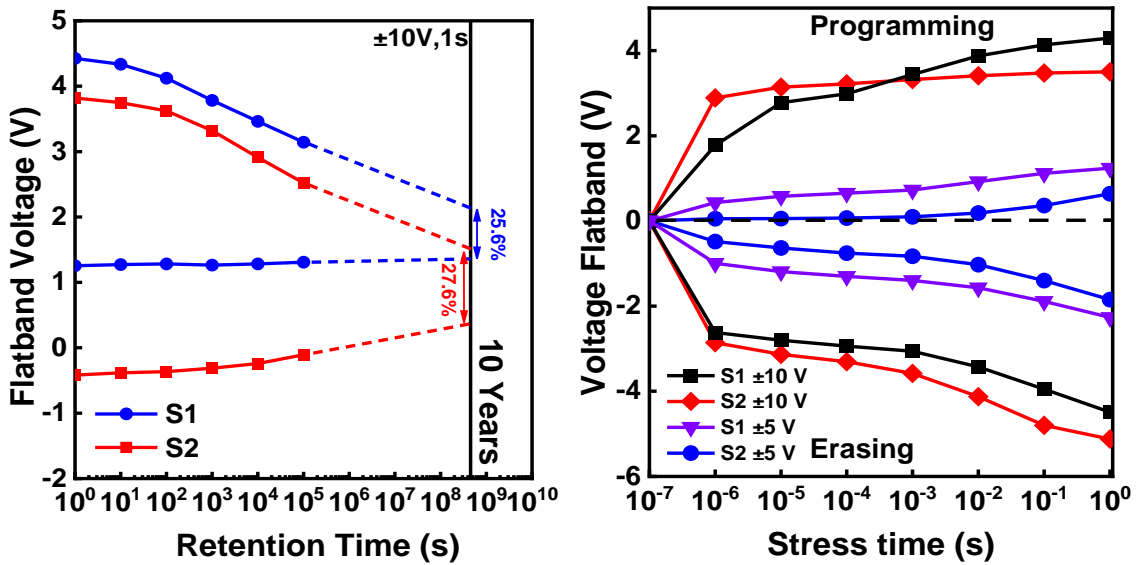

**Fig. 9. (a)** The retention properties of S1 and S2 at various P/E bias conditions **(b)** The P/E charac-

teristics for S1 and S2 at various P/E bias conditions

## 4. CONCLUSION

Various process methods have been investigated to improve the memory performance (memory window, retention performance and P/E speeds) of charge trapping memory devices. The results show that by varying the Hf:Al sputtering ratio and applying Co-60 radiation to the device, the memory window can be increased by 54%. By adjusting the tunneling layer thickness and changing the RTA treatment conditions, the retention time can be effectively extended by about 80 % and the programming and erasing speed can be increased. According to the study, most experiments focus on process conditions rather than new materials and structures, and these could be new directions for future research.

## ACKNOWLEDGEMENT

This work was supported by the National Research Foundation of Korea(NRF) grant funded by the Korea government(MSIT) (No. NRF-2022R1A4A1028702). And also, this research was funded and conducted under the Competency Development Program for Industry Specialists of the Korean Ministry of Trade, Industry and Energy (MOTIE), operated by Korea Institute for Advancement of Technology (KIAT) (No. P0012453, Next-generation Display Expert Training Project for Innovation Process and Equipment, Materials Engineers).